\documentclass[12pt]{article}
\usepackage{amsfonts, amssymb,amsmath,epsfig}
 \normalsize
\usepackage{amssymb,amsmath,amsthm,epsfig,amscd,enumerate,eucal,indentfirst}
\newtheorem{theo}{Theorem}[section]

\theoremstyle{definition}
\newtheorem{nb}[theo]{Remark}

\newenvironment{preuve}{\noindent {\tt Proof
:}}{\hfill$\blacksquare$\bigskip}
\def\a{\gamma}
\def\t{\tau}
\def\b{\theta}

\def\Q {\mathcal{Q}}

\def\QF {\mathcal{Q}_t}

\def\R{{\mathbb R}^N}
\def\Rr {{\mathbb R}^3}
\def \S {{\mathbb S}^{2}}
\def \q {\mathbf{q}}
\def \n {\mathbf{n}}
\def \qn {|\q \cdot \n|}
\def \d {\mathrm{d}}
\def \u {\mathbf{u}}
\def \v {\tilde{v}}
\def \vh {v^{\star}}

\def \w {\tilde{w}}
\def \wh {w^{\star}}
\def \wb {w_{\star}}
\def \Itt {\int_{\Rr \times \Rr \times \S}}

\def \IR {\int_{\R}}

\def\ep{\epsilon}

\def\dpi {2\pi}
\def\e2{\epsilon^2}

\def \dspl {\displaystyle}
\newcommand{\bq}{\begin{equation}}
\newcommand{\eq}{\end{equation}}
\newcommand{\bqs}{\begin{equation*}}
\newcommand{\eqs}{\end{equation*}}
\newcommand{\bqm}{\begin{multline*}}
\newcommand{\eqm}{\end{multline*}}

\def\bqa{\begin{eqnarray}}
\def\eqa{\end{eqnarray}}
\def\bd{\begin{displaymath}}
\def\ed{\end{displaymath}}

\numberwithin{equation}{section}

\title{ {\bf Long time behavior of
non--autonomous Fokker--Planck equations and the cooling of granular gases }}

\author{{\bf Bertrand Lods} \\ \normalsize Politecnico di Torino,
Dipartimento di Matematica,\\ \normalsize Corso Duca degli
Abruzzi, 24,
\normalsize10129 Torino, Italy. \\
\normalsize e-mail: {\tt lods@calvino.polito.it}\\
\\
{\bf Giuseppe Toscani} \\ \normalsize Universit\'a di Pavia,
Dipartimento di Matematica,
,\\ \normalsize Via Ferrata, 1, 27100 Pavia, Italy.\\
\normalsize e-mail: {\tt toscani@dimat.unipv.it}}

\date{}

\begin{document}

%{\bf On Fokker--Planck equations with time--dependent
%coefficients}\\

\bibliographystyle{plain}
\maketitle

\begin{abstract}
\noindent We analyze the asymptotic behavior of linear Fokker--Planck equations with
time--dependent coefficients. Relaxation towards a Maxwellian distribution with time--dependent
temperature is shown under explicitly computable conditions. We apply this result to the study of
Brownian motion in granular gases as introduced in \cite{brey}, by showing that the Homogenous
Cooling State attracts any solution at an algebraic rate.
\end{abstract}

{\small \noindent {\bf Key words.} Fokker--Planck equation,
intermediate asymptotics, granular gases, Homogeneous Cooling
State.

%\bigskip\noindent
%{\bf AMS(MOS) subject classification.} 76P05, 82C40.

%%%%%%%%%%%%%%%%%%%%%%%%%%%%%%%%%%%%%%%%%%%%%%%%%%%%%%%%%%%%%%%%%%%%%%%%%%

\section{Introduction}

The linear Fokker--Planck equation (FPE)
 \bq \label{fpconst}
  \dfrac{\partial \varphi}{\partial t}(v,t)=\lambda \nabla \cdot \left\{ v
 \varphi(v,t) + \theta \nabla \varphi(v,t)\right\}, v \in \R (N \geqslant 1)
  \eq
  arises in many
fields of applied sciences such that statistical mechanics, chemistry, mathematical finance (see
the monographs \cite{risken} and \cite{coffey} for a large account of applications to the FPE).
Such a drift-diffusion equation can be derived from the Langevin equation to model the Brownian
motion of particles in thermodynamical equilibrium. In this case, the parameters $\lambda$ and
$\theta$ are two positive constants which represent respectively the friction term and the
temperature of the system. The qualitative analysis of equation \eqref{fpconst} is
 well documented in the literature. We refer to \cite{risken} for a precise description of the hilbertian and
spectral methods used in the study of  \eqref{fpconst}  and to \cite{fokker, cartos} for the relatively recent
approach to the
$L^1$-theory by means of entropy-dissipation methods.

It is easy to notice that
the set
\bqs \label{FPE}\mathcal{M}=\left\{g \in L^1(\R),\;g \geqslant 0,\;\IR g(v) \d v=1,\;\IR v g(v)\d
v=0,\;\IR v^2 g(v) \d v=\theta < \infty \right\} \eqs
is invariant under the action of the right-hand-side of
\eqref{fpconst}. Moreover, it is well-known that \eqref{fpconst}
admits a unique steady state in $\mathcal{M}$ given by the
Gaussian distribution (\textit{Maxwellian} function in the
language of kinetic theory)
$$M_{\theta}(v)=(2\pi
\theta)^{-N/2}\exp{\left(-v^2/2\theta\right)}, \qquad v \in \R.$$ Entropy-dissipation methods
(see \cite{dolbeaut} for a review on recent results on the topic) provide a precise picture of
the asymptotic behavior of the solution to \eqref{fpconst} for initial data in $\mathcal{M}.$
Given $f \in \mathcal{M}$, the (Boltzmann)
%$H$--functional:
%\bqs H(f)=\IR f(v) \log f(v) \d v. \eqs
%and the
\textit{relative entropy} (finite or not) of $f$ is defined as
\bqs H(f\,|\,M_{f})=H(f)-H(M_{f})=\IR f \log
\left(\dfrac{f}{M_{f}}\right)\d v\eqs
where $M_{f}$ is the unique Maxwellian distribution in $\mathcal{M}$ with the same temperature as
$f$. Given $f_0 \in \mathcal{M}$, with the assumption of bounded initial relative entropy
\begin{equation*}\label{cond3}
H(f_0\,|\,M_{\theta}) \ < \ \infty ,
\end{equation*}
it has been proven in \cite{fokker} that the unique
mass-preserving solution $f(v,t)$ of \eqref{fpconst} decays
exponentially fast with rate $2\lambda$ to $M_{\theta}(v)$ in
relative entropy, i. e, the estimate
\begin{equation*}\label{decay1}
 H(f\,|\,M_{\theta})(t)\le e^{-2\lambda t}H(f_0\,|\,M_{\theta})
\end{equation*}
holds. The classical \textit{Csiszar-Kullback inequality} then
allows to translate convergence in relative entropy to the more
standard $L^1$-setting. The following result is proved in
\cite{fokker}:
\begin{theo}\label{constant}
Let $\lambda$ and $\b$ be two positive constants. Let us assume that $f_0 \in \mathcal{M}$ has a
finite relative entropy. Then,  there exists a constant $C
>0$ depending only on the initial relative entropy such that the solution
$f(v,t)$ to \eqref{fpconst} fulfills
$$\|f(\cdot,t)-M_{\theta}\|_{L^1(\R)} \leqslant C \exp{( - \lambda t)}$$
for any $t \geqslant 0$.
\end{theo}

Main objective of this paper is to generalize this result
allowing the friction term $\lambda$ and the temperature $\theta$
to fluctuate with time. In this case, the Fokker--Planck equation
reads:
\bq \label{brown}
 \dfrac{\partial \varphi}{\partial t}(v,t)=\lambda(t) \nabla
\cdot \left\{v \varphi(v,t) + \theta(t) \nabla
\varphi(v,t)\right\} \eq
where $\lambda(t)$ and $\theta(t)$ are positive functions of time.

 {\it Non--autonomous} Fokker-Planck equations arise for instance in the study
of   a periodically driven Brownian rotor \cite{alpatov} and in
this case $\lambda(t)$ and $\theta(t)$ are periodic functions of
time. In statistical mechanics, equation \eqref{brown} arises as a
natural generalization of equation \eqref{fpconst} in the context
of {\it non-equilibrium} thermodynamics \cite{luczka}. Among other
models, equation \eqref{brown} appears in the study of the tagged
particle dynamics of a heavy particle in a gas of much lighter
inelastic particles. As observed by J. J. Brey, W. Dufty and A.
Santos \cite{brey}, the large particles exhibit Brownian motion
and the Boltzmann--Lorentz kinetic equation satisfied by the
distribution function of large particles can be reduced to a
Fokker-Planck equation whose coefficients depend on the
temperature of the surrounding gas. Granular gases being
non-equilibrium systems, this temperature turns out to be
time-dependent and the Fokker-Planck equation derived in
\cite{brey} is of the shape \eqref{brown}. Since the study of the
long--time behavior of the solution to the Brey--Dufty--Santos
model is one of the main goals of our analysis, we will explain
with much more details the approach of \cite{brey} in the next
section.\medskip

Because of the time-dependence of both $\lambda(t)$ and
$\theta(t)$, equation \eqref{brown} does not possess stationary
states. Nevertheless, two natural questions arise:

\begin{itemize}

 \item[-] Do they exist particular (self-similar) solutions
to \eqref{brown} which attract all other solutions (as the
Maxwellian does in the autonomous case)?
 \item[-] If such
self-similar solutions exist, is it possible to reckon the rate
at which they attract the other solutions?
\end{itemize}

We answer positively to these two questions under some reasonable
conditions on the time--behavior of friction and temperature. Our
method is based upon suitable (time--dependent) scalings which
allow us to transform the non-autonomous equation \eqref{brown}
into a Fokker-Planck equation of the form \eqref{fpconst} and then
to make use of Theorem \ref{constant}. Clearly, the self--similar
profile is a Maxwellian with time--dependent temperature. In this
context, the Maxwellian distribution plays the role of the
Barenblatt profile in the study of the porous medium equation
\cite{vazquez}.

Application of this abstract result to our motivating example in
kinetic theory of granular gases shows that the distribution of
the Brownian particles relaxes towards a Maxwellian distribution
with time--dependent temperature. While this fact has already been
noticed by J. Javier Brey {\it et al.} \cite{brey}  our result
gives a precise \textit{estimate of the rate of convergence}
towards this self--similar profile (the so--called
\textit{Homogeneous Cooling State}) which turns out to be only
algebraic in time. Analogous results, which try to clarify the
role of the Homogeneous Cooling State in kinetic models of
granular gases, have been recently obtained for the case of the
Boltzmann equation for inelastic Maxwell particles \cite{BCT}. We
postpone a detailed discussion on this point in the conclusions of
this note.

The organization of the paper is the following. In the next section, we present in some details
the derivation of the FPE in the context of Brownian motion for granular gases. In Section 3, we
deals with a general non-autonomous FPE and we answer to the two aforementioned questions
(Theorem \ref{main}). Finally, in Section 4 we turn back to our motivated example of Brownian
particles and we show how the abstract result of Section 3 allows us to estimate the rate of
convergence towards the Homogeneous cooling state.

\section{The Brownian motion in granular gases}

The motion of heavy granular particles of mass $m$ embedded in a
low density gas whose particles have mass $m_g$ with $m_g \ll m$
has been considered in \cite{brey}. The particles under
consideration are assumed to be hard-spheres of $\mathbb{R}^3$
and, for the sake of simplicity, the diameters of the particles of
both species are assumed to be equal and normalized to unit. The
case of particles with different diameter can be investigated as
well,  and does not lead to major supplementary difficulties
\cite{brey}. The collisions between the heavy particles and the
fluid ones are partially inelastic and are characterized by a
coefficient of restitution $\epsilon \in (0,1)$. Assuming that the
concentration of heavy particles is small, one neglects the
collision phenomena between them. Let us denote by $f(\v,t)$ the
distribution function of the heavy particles having velocity $\v
\in \mathbb{R}^3$ at time $t > 0$ and by $g(\v,t)$ the
distribution function of the surrounding gas where, for
simplicity, it is assumed that these two quantities are
independent of the position. Then, the evolution of $f(\cdot,t)$
is given by the \textit{Boltzmann--Lorentz} equation, which in
weak form reads
\bq \label{eqfvt} \dfrac{\d }{\d t} \int_{\Rr}f(\v,t)\psi(\v) \d
\v=\Itt \qn f(\v,t)g(\w,t)[\psi(\vh)-\psi(\v)]\d\v\d\w\d\n\eq
%=\It \qn
%
for any test-function $\psi(\v).$ Here $\q=\v-\w$ and $(\vh,\wh)$
are the post--collisional velocities:
\bq \label{prepost} \vh=\v- \dfrac{\Delta(1+\ep)}{1+\Delta} (\q
\cdot \n) \n,\qquad \wb=\w + \dfrac{\Delta(1+\ep)}{1+\Delta} (\q
\cdot \n) \n \eq where $\Delta$  is the mass ratio $\Delta=m_g /
m$. Note that, by assumption, $\Delta \ll 1$.

To solve the linear equation \eqref{eqfvt}, one has to make explicit $g(\v,t)$. Assuming that the
binary collisions between the surrounding particles are inelastic and characterized by a constant
restitution coefficient $0 <\epsilon_g< 1,$ $g(\v,t)$ is given by a solution to the (nonlinear)
Boltzmann equation for granular hard--spheres \cite{dufty,goldh}. Leaving details to the
pertinent literature, the role of the equilibrium Maxwellian function in the elastic Boltzmann
equation  is here represented by the {\it Homogeneous Cooling State} (see References
\cite{ernst,goldh}) which implies that
 $$
 g(\v,t)=v_g(t)^{-3}\Phi\left(\dfrac{\v}{v_g(t)}\right)
 $$
where $v_g(t)$ is the thermal velocity of the gas particles
defined as $v_g(t)=\left[2T_g(t)/m_g\right]^{1/2}$. The
temperature $T_g(t)$ is defined in a standard way (see Section 3).
The self--similar profile $\Phi(\cdot)$ is a stationary solution
of some suitable steady Boltzmann equation (see \cite{ernst,
dufty}) and is not explicitly known. However, an important fact to
be noticed is that $\Phi(\v)$ is a function of $\epsilon_g$ which,
in the quasi--elastic regime $\epsilon_g \to 1$, converges toward
the Maxwellian distribution $\pi^{-3/2}\exp{(-\v^2)}.$ The
temperature $T_g(t)$ is cooling because of the inelasticity of the
collisions. Hereafter, we will assume $T_g(t)$ to obey the the
so-called \textit{Haff's law} \cite{haff}: \bq \label{hafflaw}
T_g(t)=T_g(0)\left(1+ t/\tau_0\right)^{-2}\eq where $\tau_0
> 0$ is the characteristic time \cite{brey}:
\bq \label{tau0} \tau_0^{-1}=\dfrac{\pi (1-\epsilon_g^2)}{12}
\sqrt{\dfrac{T_g(0)}{2m_g}} \int_{\Rr \times \Rr}\Phi(\v)\Phi(\w)
|\v-\w|^3 \d \v \d \w.\eq
%
%\begin{nb} We adopt here a general Haff's law (i.e with arbitrary positive $\nu$) even if,
%for the case under consideration here of constant coefficient
%$\epsilon_g$, Haff proved \cite{haff} that $\nu=2.$ We choose here
%an arbitrary positive $\nu$ only in a pedagogical perspective
%since we will see hereafter that Theorem \ref{main} applied if and
%only if $\nu=2$. This is a kind of {\it a posteriori} proof of
%Haff's law. We recall also that $\nu=5/3$ for a gas of
%viscoelastic spheres \cite{schwager}\end{nb}
%

The Boltzmann-Lorentz equation \eqref{eqfvt} can be reduced to a
FPE with time--dependent coefficients of the form \eqref{brown}
performing two asymptotic procedures:\medskip

 $\bullet$ The first procedure is a simple extension of
the standard method for elastic particles (grazing collisions
asymptotics \cite{cerci}, see also \cite{BLGT} in the context of
the dissipative linear Boltzmann equation). Precisely, according
to \eqref{prepost}, one sees that, when a heavy particle collides
with a small one, the velocity of the heavy particle is only
slightly altered:
$$|\vh-\v|=\left|\dfrac{\Delta(1+\ep)}{1+\Delta} (\q \cdot \n)
\n\right| \ll 1$$ so that $\vh \simeq \v$. Therefore, performing
in \eqref{eqfvt} a formal expansion to leading order in the mass
ratio as $\Delta \to 0$, one obtains the following Fokker--Planck
equation with time--dependent coefficients:
\bq \label{eqFPfvt} \dfrac{\partial f(\v,t)}{\partial t} =
 \nabla_{\v} \cdot \left[\mathbf{A}(\v,t)
f(\v,t) + \dfrac{1}{2} \nabla_{\v} \cdot \left(
\mathbb{N}(\v,t)f(\v,t) \right) \right]\eq
where the vector $\mathbf{A}(\v,t)$ and, respectively, the tensor $\mathbb{N}(\v,t)$ are given by
 $$
 \mathbf{A}(\v,t)=\dfrac{(1+\ep)\Delta}{1+\Delta}\dfrac{\pi}{2}\int_{\Rr}g(\w,t)
\q |\q| \d \w,
 $$
 and
 $$
 \mathbb{N}_{ij}(\v,t)=\left(\dfrac{1+\epsilon \Delta}{1+\Delta}\right)^2\dfrac{\pi}{12}
\int_{\Rr}g(\w,t) \left(|\q|^3\delta_{ij} + 3 |\q|\q_i\q_j
\,\right) \d \w \qquad  (i,j=1,2,3).
 $$
 We used above the standard
notation
 $$
 \nabla_{\v} \cdot (\mathbb{N}(\v)
f(\v))=\left(\displaystyle \sum_{j=1}^3\dfrac{\partial }{\partial
\v_j}\mathbb{N}_{ij}(\v,t)f(\v,t)\right)_{i=1,2,3}
 $$
 (see
\cite[Appendix A]{brey} for a detailed derivation).\medskip

$\bullet$ To simplify further the Fokker-Planck equation \eqref{eqFPfvt} one performs a
second asymptotic procedure  which consists in assuming that the thermal
velocity of the heavy particles $v(t)=\left[2T(t)/m\right]^{1/2}$
is negligible with respect to the one of the surrounding particles
$v_g(t)$: $v(t) \ll v_g(t).$ This leads to a formal expansion in
$T(t)\Delta/ T_g(t)$  where $T(t)$ is the temperature of the
Brownian particles (see Remark \ref{ratio}). In this case, the
vector $\mathbf{A}(\v,t)$ and the tensor $\mathbb{N}(\v,t)$ reduce
to
$$\mathbf{A}(\v,t)\simeq \alpha(t) \v \qquad \text{ and }
\qquad \mathbb{N}_{ij}(\v,t) \simeq 2 \eta(t) \delta_{ij}, \qquad
i,j= 1,2,3$$ where
\bq \label{alphaeta} \alpha(t)=\zeta T_g(t)^{1/2} \qquad \text{
and } \qquad \eta(t)=\xi T_g(t)^{3/2}. \eq
In    \eqref{alphaeta} we used the notations
\bq \label{zeta}
\zeta=\dfrac{2\sqrt{2}\pi}{3\sqrt{m_g}}(1+\epsilon)\Delta
\int_{\Rr}\Phi(\w)|\w|\d \w  ,
 \eq
and
$$
\xi=\dfrac{\pi m_g^{-3/2}}{3\sqrt{2}
}(1+\epsilon)^2\Delta^{2}\int_{\Rr}\Phi(\w)|\w|^3\d \w.
 $$
 Taking
these simplifications into account, equation \eqref{eqFPfvt} is
replaced by the following
\bq \label{FPbrown}
\dfrac{\partial f}{\partial t}(\v,t)=  \nabla_{\v} \cdot
\displaystyle \left\{ \alpha(t) \v \, f(\v,t) + \eta(t)
\nabla_{\v} f(\v,t) \right\}.
 \eq
The Fokker-Planck equation \eqref{FPbrown}  is of the form
\eqref{brown}. The drift and diffusion coefficients depend on time
only through the surrounding gas temperature $T_g(t)$ (see
\eqref{alphaeta}).

\begin{nb}\label{ratio} The derivation of \eqref{FPbrown} from \eqref{eqfvt} is based upon a \textit{time--dependent} asymptotic
procedure where it is assumed that \bq \label{validtemp}
\frac{T(t)\Delta}{T_g(t)}\ll 1. \eq
As shown in \cite{brey} this assumption
requires  $\Delta \to 0$ and $\epsilon_g \to 1$
(quasi--elastic regime for the surrounding gas) simultaneously. A further consequence of
these assumptions is that
\begin{equation}\label{limi}
\dfrac{1}{2\sqrt{2}\Delta} \dfrac{1-\epsilon_g^2}{1+\epsilon}<
1.\end{equation}
We will find again this condition hereafter.
\end{nb}

\section{Long time behavior of non--autonomous Fokker--Planck equations}

In this section we consider the general Fokker--Planck collision
operator with \textit{time--dependent coefficients} written in the
divergence form: \bq \label{FPT}
 \QF (f)(v)= \lambda(t)  \nabla_{v} \cdot \left \{ v \, f(v,t) + \theta(t) \nabla_{v} f(v,t) \right\},
\quad
 v \in \R (N \geqslant 1)  \eq
where $\lambda(t)$ and $\theta(t)$ are two positive functions of
time.  We are concerned with the large--time asymptotic behavior
of the solution to the Cauchy problem
\bq \label{CP}
\begin{cases} \dfrac{\partial f}{\partial
t}(v,t)=\QF(f)(v) \qquad v \in \R, t > 0\\
f(v,0)=f_0(v) \end{cases}\eq where the initial data $f_0$ is
assumed to be nonnegative and integrable,
$$f_0 \geqslant 0, \text{ and } f_0 \in L^1(\R).$$
In accordance with the language of kinetic theory, we define the
mass density $\varrho(t)$, mean velocity $\u(t)$ and temperature
$T(t)$ respectively as:
$$\varrho(t)=\IR f(v,t)\d v, \qquad \u(t)=\dfrac{1}{\varrho(t)}\IR
vf(v,t)\d v,$$ and
$$T(t)=\dfrac{1}{N\varrho(t)}\IR |v-\u(t)|^2f(v,t)\d v.$$
The number density is preserved by the (non--autonomous)
Fokker--Planck operator \eqref{FPT}  while the mean velocity is
preserved only if initially equal to zero. Precisely, if $\IR
vf_0(v)\d v=0,$ then $\u(t)=0$ for any $t > 0.$ In this case, the
evolution of the temperature is \bq \label{evoltemp} \dfrac{\d
T(t)}{\d t}=-2 \lambda(t) \left(T(t)-\b(t)\right), \qquad \qquad
(t > 0).\eq
%
%Clearly, the set $\mathcal{M}$ is left invariant by the collision
%operator $\QF$.
In order to find the intermediate asymptotic for \eqref{CP}, we
look for a solution to \eqref{FPT} of the shape:
$$f(v,t)=\a(t)^{-N}F\left(v/\a(t),\tau(t)\right)=\a(t)^{-N}F(\v,\tau)$$
where the new time scale $\tau=\tau(t)$ is nonnegative and such
that $\tau(0)=0$,  the scaled velocity is
$$\v=v/\a(t)$$ and
$\a(\cdot)$ is positive. Without loss of generality, one may
assume that $\a^2(0)=T_0:=\int_{\Rr}v^2 f_0(v)\d v$ so that
$$F(\v,0)=F(v/T_0,0)=f_0(v/T_0).$$
One sees immediately that
\bq \label{dft} \dfrac{\partial f}{\partial
t}(v,t)=\dfrac{\dot{\tau}(t)}{\a(t)^N} \dfrac{\partial F}{\partial
\tau}(\v,\tau) - \dfrac{\dot{\a}(t)}{\a(t)^{N+1}} \nabla_{\v}
\cdot \left(\v F(\v,\tau) \right), \eq
where the  dot symbol stands for the time derivative. In the same
way, one can show that
\bq \label{qf}\Q_t(f)(v)=
\dfrac{\lambda(t)}{\a(t)^N}\nabla_{\v} \cdot (\v F(\v,\tau))+
\dfrac{\lambda(t)\b(t)}{\a(t)^{N+2}} \nabla_{\v}^2 F(\v,\t). \eq
This leads to the following evolution equation for $F(\cdot, \t)$:
\begin{equation}\label{Ftau}
\dfrac{\partial F}{\partial \t}(\v,\t)=\dfrac{1}{\dot{\t}(t)}
\left[ \lambda(t) + \dfrac{\dot{\a}(t)}{\a(t)} \right] \nabla_{\v}
\cdot (\v
F(\v,\tau))\\
+ \dfrac{\lambda(t)\b(t)}{\dot{\t}(t) \, \a(t)^2} \nabla_{\v}^2
F(\v,\t) \end{equation}
One notes that \eqref{Ftau} reduces to a {\it ''good''}
Fokker--Planck equation
\begin{equation}\label{dtau}\dfrac{\partial F}{\partial \t}(\v,\t)= \nabla_{\v} \cdot \left(
 \v \,F(\v,\t) + \nabla_{\v} F(\v,\t) \right) \qquad \v \in
\R, \, \t
> 0\end{equation}
provided %there exists some $\sigma > 0$ such that
\begin{equation}\label{hyp1}
\dfrac{1}{\dot{\t}(t)} \left[ \lambda(t) +
\dfrac{\dot{\a}(t)}{\a(t)} \right]=1 \qquad \forall t > 0,
\end{equation}
and
\begin{equation}\label{hyp2}
\dfrac{\lambda(t)\b(t)}{\dot{\t}(t) \, \a(t)^2}=1 \qquad \qquad
\forall t > 0.
\end{equation}
Of course, to investigate the asymptotic behavior of $F(\cdot,\t)$
and apply Theorem \ref{constant}, one has to find conditions on
$\lambda(\cdot)$ and $\b(\cdot)$ insuring that the time scale $\t$
verifies
$$\lim_{t \to \infty} \t(t)=+\infty.$$
Solving equation \eqref{hyp1}--\eqref{hyp2} leads to
$$\dfrac{\lambda(t) \b(t)}{\a^2(t)}=\dot{\t}(t) =
\lambda(t) + \frac{\dot{\a}(t)}{\a(t)}$$ i. e.
$$\lambda(t) \theta(t) = \a^2(t) \lambda(t) + \a(t)
\dot{\a}(t) = \a^2(t)\lambda(t)+ \dfrac{1}{2}\dfrac{\d}{\d
t}\{\a^2(t)\}.$$ Since $\a(0)=\sqrt{T_0}$, one obtains
\begin{equation}\label{gamma}
\gamma(t)=\exp{\left(-\int_0^t\lambda(s)\d s\right)}\left\{ T_0 + 2\int_0^t
\lambda(s)\b(s) \exp{\left(2 \int_0^s \lambda(r) \d r\right)} \,  \d s
\right\}^{1/2} \qquad t > 0.\end{equation}
Now, from \eqref{hyp2},
\begin{equation*}\begin{split}
\dot{\t}(t)&= \dfrac{\lambda(t)
\theta(t)}{\a^2(t)}=\dfrac{\lambda(t)\b(t)\,\exp{\left( 2  \dspl
\int_0^t\lambda(s)\d s\right)}}{ T_0 + 2 \dspl \int_0^t
\lambda(s)\b(s) \exp{\left(2 \dspl \int_0^s \lambda(r) \d r\right)} \, \d s}\\
&= \dfrac{1}{2} \dfrac{\d}{\d t} \log{\left [T_0+ 2 \dspl \int_0^t
\lambda(s)\b(s) \exp{\left(2 \dspl \int_0^s \lambda(r) \d r\right)}\,\d s
\right]}.\end{split}\end{equation*}
Solving this equation with the initial datum $\tau(0)=0$ one
gets
\begin{equation}\label{tau}
\tau(t)=\dfrac{1}{2}\log{\left[T_0 + 2 \dspl \int_0^t
\lambda(s)\b(s) \exp{\left(2 \dspl \int_0^s \lambda(r) \d r\right)}\, \d
s\right]} \qquad \qquad (t > 0).\end{equation} Clearly,
$$\lim_{t \to \infty} \tau(t)=+\infty \quad \text{if and only if}
\quad \int_0^{\infty}\lambda(t)\b(t)\exp{\left(2 \dspl \int_0^t
\lambda(s) \d s\right)}\,\d t = \infty.$$
This leads to the following result.
\begin{theo}\label{main}
Let us assume that $\lambda(\cdot)$ and $\b(\cdot)$ are
nonnegative functions on $\mathbb{R}_+$ satisfying
\begin{equation}\label{infini}
\int_0^{\infty}\lambda(t)\b(t)\exp{\left(2 \dspl \int_0^t
\lambda(s) \d s\right)}\,\d t = \infty.\end{equation}
Let us assume furthermore that $f_0 \in \mathcal{M}$ has a finite relative
entropy. Then, there exists a constant $C
> 0$ such that
 \begin{equation}
\|f(\cdot,t)-f_{\infty}(\cdot,t)\|_{L^1(\R_v)} \leqslant
\dfrac{C}{ \left \{ T_0+2 \dspl \int_0^t
\lambda(s)\b(s)\exp{\left(2\int_0^s \lambda(r) \d r\right)} \,\d
s\right\}^{1/2} } \qquad t > 0.
 \end{equation}
 The
intermediate asymptotic profile $f_{\infty}(v,t)$ is given by
 $$
  f_{\infty}(v,t)=(\dpi \, T(t))^{-N/2} \exp{ \{ -v^2 /
2 \,T(t)\}}=M_{T(t)}(v)
 $$
 where $T(t)$ is the temperature of
$f(\cdot, t)$ given by \eqref{gamma}.
\end{theo}
\begin{preuve} The proof reduces to the study of \eqref{Ftau}. Clearly, one may choose $\sigma=1$ (this is
equivalent to change $\tau$ by \eqref{tau}). By \eqref{infini},
 $\t(t) \to \infty$, and according to Theorem \ref{constant},
\begin{equation}\label{esttau}
\|F(\cdot,\t)-M_1(\cdot)\|_{L^1(\R_{\v})} \leqslant C \exp{\{-
\tau \}} \qquad \qquad \tau > 0\end{equation} provided $F(\v,0)$
is of finite relative entropy. This is the case since
$F(\v,0)=f_0(v/T_0)$. Now, turning back to the original variables,
one gets the conclusion using the fact that
$$\exp\{-\tau(t)\}=\left\{1 + 2 \dspl \int_0^t
\lambda(s)\b(s) \exp{\left(2 \dspl \int_0^s \lambda(r) \d r\right)}\, \d
s\right\}^{-1/2} \qquad \qquad (t > 0)$$ by virtue of \eqref{tau}.
\end{preuve}

\begin{nb} The intermediate asymptotic is given by the Maxwellian distribution with the same
temperature $T(t)$ as the one of $f(\cdot,t)$. Of course, this
Maxwellian distribution is a particular solution to \eqref{CP}.
% We point out that, according to \eqref{evoltemp}, the temperature $T(t)$
% coincide with $\gamma^2(t)$ (see \eqref{alphaeta}) with $\sigma=T(0)^{-1}.$\end{nb}
The most important feature of Theorem \ref{main} is that it
provides the rate of convergence of any solution $f(v,t)$ towards
the self-similar profile $f_{\infty}(v,t)$. This rate is explicit
in terms of the known coefficients $\lambda(t)$ and
$\theta(t)$.\end{nb}

\section{The homogeneous cooling state for the Brownian particles}

We apply here the results of Section 3 to the study of the
so--called {\it homogeneous cooling state} for equation
\eqref{FPbrown}. Let $f_0(\v)$ be an element of $\mathcal{M}$ %with unit temperature $\int_{\Rr} \v^2 f_0(\v)\d \v=1$
 and let us
consider the Cauchy problem
\bq \label{Cauchy}
\begin{cases} \dfrac{\partial f}{\partial
t}(\v,t)=\lambda(t)  \nabla_{\v} \cdot \left \{ \v \, f(\v,t) +
\theta(t) \nabla_{\v} f(\v,t) \right\}
 \qquad \v \in \Rr, t > 0\\
f(\v,0)=f_0(\v) \end{cases}\eq where we transformed the
right--hand side of equation \eqref{FPbrown} into a non--autonomous
Fokker--Planck operator of the form \eqref{FPT} by setting,
according to \eqref{alphaeta},
\bq \label{lamb} \lambda(t)=\alpha(t)= \zeta T_g(t)^{1/2} \qquad
\text{ and } \qquad \theta(t)=\dfrac{\eta(t)}{\alpha(t)}=\xi
\zeta^{-1} T_g(t).\eq
Accordingly, for any $t > 0$ and with the notations of Section 3
\bqs
\begin{split} \int_0^t \lambda(s)\theta(s)&\exp{\left(2 \int_0^s
\lambda(r)\d r\right)} \d s=\xi \int_0^t T_g(s)^{3/2} \exp\left( 2
\zeta \int_0^s
\sqrt{T_g(r)} \d r\right) \d s\\
&=\xi T_g(0)^{3/2}\int_0^{t} (1+s/\tau_0)^{-3} \exp{\left( 2 \zeta
\sqrt{T_g(0)} \int^s_0 \frac{\d r}{1+r/\tau_0}\right)} \d s .
\end{split}
\eqs
Note that the last equality follows from Haff's law \eqref{hafflaw}. Clearly, for any $s
> 0$
$$(1+s/\tau_0)^{-3} \exp{\left( 2 \zeta
\sqrt{T_g(0)}\int^s_0 \frac{\d
r}{1+r/\tau_0}\right)}=(1+s/\tau_0)^\nu$$ with $\nu=2\zeta \tau_0
\sqrt{T_g(0)}-3.$ Consequently, condition \eqref{infini} of
Theorem \ref{main} is verified provided $\nu \geqslant -1.$
Note however that, if $\nu=-1$, then \bqs
\begin{split}
T(t)&=\exp{\left(-2\int_0^t\lambda(s)\d s\right)}\left\{ T_0 + 2\int_0^t
\lambda(s)\b(s) \exp{\left(2 \int_0^s \lambda(r) \d r\right)} \,  \d s
\right\}\\
&=(1+t/\tau_0)^{-2}\left(T_0+\xi T_g(0)^{3/2} \tau_0
\log(1+t/\tau_0)\right),\end{split}\eqs
so that
$$\dfrac{T(t)}{T_g(t)} \longrightarrow \infty \qquad \text{ as } \qquad  t \to \infty.$$
In this case, assumption \eqref{validtemp} is violated, and the
Fokker-Planck equation \eqref{FPbrown} is meaningless. Hence,
condition \eqref{infini} of Theorem \ref{main} reduces to $\nu >
-1$, i. e. \bq \label{zettau} \zeta \tau_0 \sqrt{T_g(0)} > 1.\eq
Using \eqref{tau0} and \eqref{zeta} the above condition reads
$$16\dfrac{(1+\ep)\Delta}{1-\epsilon_g^2}\dfrac{%\displaystyle
\int \Phi(\v)|\v|\d \v} {%\displaystyle
\int \Phi(\v)\Phi(\w)|\v-\w|^3 \d \v \d \w} > 1.$$
Now, recall that the Fokker-Planck equation \eqref{FPT} turns out to be valid only
for nearly elastic surrounding particles (see Remark \ref{ratio}).
Since in this quasi--elastic regime $\Phi(\v)$ approaches the
Maxwellian distribution $\pi^{-3/2} \exp(-\v^2)$, one can reasonably
approximate the two moments $\int_{\Rr}\Phi(\v)|\v|\d \v$ and
$\int_{\Rr \times \Rr} \Phi(\v)\Phi(\w)|\v-\w|^3 \d \v \d \w$ by
their limits as $\epsilon_g \to 1$, obtaining
$$\int_{\Rr} \pi^{-3/2} \exp{\left(-\v^2\right)} |\v| \d
\v= 2/\sqrt{\pi}$$ and $$\int_{\Rr \times \Rr}
\pi^{-3}\exp{\left(-\v^2-\w^2\right)}|\v-\w|^3\d \v \d
\w=16/\sqrt{2\pi}.$$
Using these approximation, equation \eqref{zettau} turns out to be
equivalent to
$$2\sqrt{2}\Delta
\dfrac{1+\ep}{1-\ep_g^2} > 1.$$
Once again, we find the same condition \eqref{limi} of validity of the
non--autonomous Fokker--Planck equation \eqref{FPbrown}.

\begin{nb} Note that, here again, we only assume that the
surrounding gas particles suffer  nearly elastic collisions (i.e.
$1-\ep_g \ll 1$) but we not  assume $\ep_g$ to be equal to one. As a
consequence, we do not replace the cooling state  profile  $\Phi$ by the
Maxwellian distribution, but we only assume that its moments do not
differ to much from the ones of the Maxwellian distribution.
\end{nb}
The previous reasoning leads to the following Theorem.
\begin{theo}\label{granul}
Let us assume that
$$\dfrac{1}{2\sqrt{2}\Delta}\dfrac{1-\ep_g^2}{1+\ep} \longrightarrow \beta^{-1} <
1   \qquad \text{ as } \qquad \Delta \to 0,\; \ep_g \to 1.$$ Let
$f_0 \in \mathcal{M}$ be of finite relative entropy. Then, the
solution $f(\v,t)$ to the Fokker-Planck equation \eqref{FPT}
converges towards the {\it cooling} Maxwellian
$$f_{\infty}(\v,t)=(2\pi T(t))^{-3/2} \exp\left\{-\v^2 /
2T(t) \right\},
 $$
and the following bound holds
 $$\|f(\cdot,t)-f_{\infty}(\cdot,t)\|_{L^1(\Rr)}=O(t^{1-\beta})
\qquad (t \to \infty).
 $$
 The temperature of the Maxwellian is given by
$$T(t)=\dfrac{(1+\ep)\Delta T_g(0)}{2m_g(1-\beta^{-1})}
\left(1+t/\tau_0\right)^{-2} + \left(T(0)-\dfrac{(1+\ep)\Delta
T_g(0)}{2m_g(1-\beta^{-1})} \right)
\left(1+t/\tau_0\right)^{-2\beta}.$$
\end{theo}
\begin{preuve} The proof is a straightforward application of
Theorem \ref{main}. Here we use the fact that \bqs
\begin{split} \int_0^t \lambda(s)\theta(s)\exp{\left(2 \int_0^s
\lambda(r)\d r\right)} \d s&=\xi T_g(0)^{3/2} \int_0^t
(1+s/\tau_0)^{2\beta - 3} \d s\\
&=\dfrac{\tau_0}{2\beta-2}\xi T_g(0)^{3/2} \left\{
(1+t/\tau_0)^{2\beta-2}-1\right\}.\end{split}\eqs
Moreover, by \eqref{gamma}, the temperature of the heavy particles
is
$$T(t)=\exp \left(-2\int_0^t
\lambda(s) \d s\right) \left\{T(0)+2\int_0^t
\lambda(s)\theta(s)\exp{\left(2 \int_0^s \lambda(r)\d r\right)} \d
s \right \}.$$ This last quantity is explicitly computable using
the expressions of $\tau_0$ and $\xi$.
\end{preuve}

\begin{nb} One notes that $T(t)$ obeys asymptotically Haff's law since the decay of
temperature of the heavy particles for large $t$ is in
$O\left((1+t/\tau_0)^{-2}\right)$. Actually,
$$\dfrac{T(t)}{T_g(t)} \longrightarrow \dfrac{(1+\ep)\Delta}{2m_g(1-\beta^{-1})} \qquad \text{ as } \qquad t \to \infty.$$
An interesting feature is that, depending on the values of the
parameters $\ep,$ $\Delta$, $m_g$ and $\beta$, the temperature of
the heavy particles is greater or smaller than the one of the
surrounding gas. This contrasts the classical case of elastic
particles in equilibrium. Indeed, in this case, according to
Theorem \ref{constant}, the distribution function relaxes to a
Maxwellian distribution whose temperature $\theta$ is exactly the
one of the surrounding bath (see \cite{fokker} for more details).
We refer the reader to \cite{brey} for a discussion of the
competing effects which imply the asymptotic difference between
$T(t)$ and $T_g(t)$.
\end{nb}

\begin{nb} We point out that, whereas for systems in equilibrium,
the relaxation rate is exponential (Theorem \ref{constant}), one
notes here that the {\it Homogeneous Cooling State}
$f_{\infty}(\v,t)$ attracts the distribution function $f(\v,t)$
only with an algebraic rate.\end{nb}

%\begin{nb} An important point to notice is that, though we
%assumed the surrounding gas to be  in its quasi--elastic regime,
%we {\it did not} replace the (unknown) distribution function
%$\Phi(\cdot)$ by the Maxwellian distribution corresponding to
%$\ep_g=1.$ \end{nb}

\section{Concluding remarks}

We discussed in this paper the intermediate asymptotics of a
linear Fokker--Planck equation with time--dependent coefficients
of the form \eqref{brown}. We showed that, under some reasonable
assumptions on the drift and diffusion coefficients, any solution
$f(v,t)$ to \eqref{brown} relaxes towards a Maxwellian distribution function
whose (time--dependent) temperature is the one of $f(v,t)$. More
important is the fact that the rate of convergence towards this
self--similar solution is explicitly computable in terms of the
coefficients and the initial temperature.

We applied our result to the motivating example of Brownian motion in granular fluids, already addressed in
\cite{brey}. For such a model, the Fokker-Planck equation \eqref{brown} is an approximation of the
Boltzmann--Lorentz equation. According to our general result (Theorem \ref{main}), the so--called Homogeneous
Cooling State for this model is a Maxwellian distribution whose temperature obeys asymptotically the Haff's law.
Moreover, the rate of convergence towards this self-similar solution is algebraic in time. We wish to emphasize
here the fact that the question of the rate of convergence towards the Homogenous Cooling State (HCS) is of primary
importance in the kinetic theory of gases. We recall here that any solution to the non--linear Boltzmann equation
for inelastic interactions relaxes towards a Dirac mass because of the dissipation of the kinetic energy.  It has
been conjectured however by Ernst and Brito \cite{ernst, brito} that the HCS attracts any solution faster than the
Dirac mass does. While for hard--spheres interactions, only few results support this conjecture, the situation is
almost clear for the non realistic case of the Boltzmann equation for maxwellian molecules \cite{BCT}. For this
model in fact, it has been shown that the (unique) self--similar solution attracts any other solution corresponding
to an initial density with a sufficiently high number of  bounded moments. Other simplified models have been
studied recently. For nearly elastic flows in one--dimension, the nonlinear Boltzmann equation reduces to a
nonlinear friction equation \cite{namara,toscani} and it has been shown in \cite{villani} that, in this case, the
HCS does not attract much faster than the Dirac mass since the improvement in the rate of convergence is only
logarithmic in time. This question has also been addressed recently in \cite{hail} for general nonlinear friction
equations corresponding to a relative velocity dependent coefficient of restitution.

\bigskip

\noindent {\bf Acknowledgement:} Part of this research was carried
out during the stay of B. Lods at the IMATI--CNR/Department of
Mathematics of the University of Pavia. He would like to express
his sincere gratitude to P. Pietra and G. Toscani for their kind
hospitality during this period. The work of B. L. at Pavia was
supported by the IHP project ``HYperbolic and Kinetic Equations'',
No.~HPRN-CT-2002-00282, funded by the EC. G. T. acknowledges
financial supports both from the project ``HYperbolic and Kinetic
Equations'', funded by the EC., and from the Italian MURST,
project ``Mathematical Problems in Kinetic Theories''.

{\small
\bibliographystyle{plain}

}
\end{document}